# Spin heat accumulation and spin-dependent temperatures in nanopillar spin valves

F.K. Dejene[1†]*, J. Flipse[1†], G. E. W. Bauer[2,3,4] & B. J. van Wees[1]

[1]*Zernike Institute for Advanced Materials, Physics of Nanodevices, University of Groningen, 9747 AG Groningen, The Netherlands*

[2]*Kavli Institute of NanoScience, Delft University of Technology, 2628 CJ Delft, The Netherlands*

[3]*Institute for Materials Research, Tohoku University, Sendai 980-8577, Japan*

[4]*WPI Advanced Institute for Materials Research, Tohoku University, Sendai 980-8577, Japan*

[†]These authors contributed equally to this work

* F.K.Dejene@rug.nl

**Since the discovery of the giant magnetoresistance (GMR) effect[1,2] the use of the intrinsic angular momentum of the electrons has opened up new spin based device concepts. The two channel model of spin-up and spin-down electrons with spin-dependent conductivities very well describes spin and charge transport in such devices. In studies of the interaction between heat and spin transport, or spin caloritronics, until recently it was assumed that both spin species are always at the same temperature. Here we report the observation of different temperatures for the spin up ($T_\uparrow$) and spin down ($T_\downarrow$) electrons in a nanopillar spin valve subject to a heat current. The weak relaxation, especially at room temperature, of the spin heat accumulation ($T_s = T_\uparrow - T_\downarrow$) is essential for its detection in our devices. Using 3D finite element modeling[3] spin heat accumulation (SHA) values of 120 mK and 350 mK are extracted at room temperature and 77 K, respectively, which is of the order of 10% of the total temperature bias over the pillar. This technique uniquely allows the study of inelastic spin scattering at low energies and elevated temperatures, which is not possible by spectroscopic methods.**

Most recent work in spin caloritronics[4,5] aimed at spin-dependent thermoelectric effects led to the discovery of thermally driven spin sources [6-10], cooling/heating by spin currents [11,12], the magneto Seebeck[13-15] and Seebeck rectification[16] in magnetic tunnel junctions. Hatami et al.[17] predicted spin-dependent temperatures in spin valve structures for sufficiently weak inter-spin heat exchange. The spin heat relaxation by inelastic scattering leads to a breakdown of the Wiedemann-Franz relation[18] between the charge and electronic heat conductance of the spin valve[17,19].

A spin-dependent temperature builds up in spin valves when the thermal conductivity ($\kappa$) in the ferromagnet ($\kappa_\uparrow \neq \kappa_\downarrow$) is spin polarized and the spin flip and inelastic scattering is sufficiently weak[17,19]. The Wiedemann-Franz relation tells us that the electronic part of the heat conductance in metals ($\kappa_e$) is proportional to the electrical conductivity ($\sigma$), with a polarization $P_\kappa = \frac{\kappa_{e\uparrow} - \kappa_{e\downarrow}}{\kappa_e}$ that should then be equal to $P_\sigma = \frac{\sigma_\uparrow - \sigma_\downarrow}{\sigma}$. A heat current through a ferromagnetic metal (F) will therefore be spin polarized, creating a spin heat accumulation (SHA) by the spin-heat coupling at an interface with a non-magnetic metal (N) (see fig. 1). If there would be no inelastic scattering of the electrons this SHA decays with the same spin relaxation length ($\lambda_s$) as the spin accumulation, i.e. the difference in the local chemical potential of the spin species. In real physical systems though, the ever-present inelastic phonon and electron-electron scattering leads to the exchange of heat between the two spin channels thereby equilibrating $T_\uparrow$ and $T_\downarrow$ to the same average temperature (fig. 1). Spin temperatures equilibrate over the *spin heat relaxation length* ($\lambda_Q$) which is either limited by spin flip scattering ($\lambda_Q = \lambda_s$) or by inelastic scattering (when $\lambda_Q < \lambda_s$). The thermal equivalent for the diffusion equation for the spin accumulation reads:

$$\nabla^2 T_s = T_s/\lambda_Q^2 \qquad (1)$$

where $T_s = T_\uparrow - T_\downarrow$ is the SHA. The temperature drop that builds up at the F/N interface (see Fig. 1) then becomes (see Supplementary A):

$$\Delta T = \frac{1}{2} \cdot P_\kappa \cdot T_s \qquad (2)$$

In regular current perpendicular to plane (CPP) spin valve devices inelastic scattering is caused by electron-phonon (e-ph) and electron-electron (e-e) interactions[19]. Time-domain thermoreflectance and BEEM studies[20,21] on inelastic scattering of hot electrons in copper found an inelastic (charge) equilibration length of the order of 60 nm, which is more than five times smaller than $\lambda_s = 350$ nm at room temperature[20]. As long as the copper spacer layer in a spin valve is comparable to $\lambda_Q$ the SHA should be detectable by the second ferromagnetic layer. In Fig. 2, $T_\uparrow$ and $T_\downarrow$ are plotted for the parallel (P) and antiparallel (AP) alignment of the magnetic layers in such a CPP spin valve. For P the SHAs at both F/N interfaces have opposite sign and sum up to be negligibly small. On the other hand, in the AP configuration both interfaces contribute constructively to generate a large SHA leading to a significant temperature drop $\Delta T$ (see Eq. (2)) at both F/N interfaces that should be observable at the bottom of the pillar. If $\lambda_Q = \lambda_s$ the Wiedemann-Franz relation holds, i.e. the relative thermal conductance ratio $\left(\frac{\kappa_P - \kappa_{AP}}{\kappa_p}\right)$ equals the GMR ratio $\left(\frac{\sigma_P - \sigma_{AP}}{\sigma_p}\right)$. However, in the presence of inter-spin and spin-conserving inelastic scattering $\lambda_Q < \lambda_s$ and we may expect that the Wiedemann-Franz relation to break down, since heat exchange short-circuits the spin channels, thereby decreasing $\kappa_P - \kappa_{AP}$ but not $\sigma_P - \sigma_{AP}$.

To observe the SHA, we use a nanopillar spin valve ($Ni_{80}Fe_{20}$/Cu/ $Ni_{80}Fe_{20}$ stack with dimensions 150×80 nm² and a thickness of each layer of 15 nm) as shown in fig. 3. We measure the temperature of the bottom contact using a Pt-Constantan ($Ni_{45}Cu_{55}$) thermocouple (contacts 3 and 4) while sending a charge current through the Pt-heater (contact 1 to 2). Both the thermocouple and the heater are electrically isolated from the bottom contact by an $Al_2O_3$ barrier (~ 8 nm thick). All samples were initially characterized by electrical measurements of the four-probe electrical resistance of the nanopillar using contacts 6 and 8 while sending a charge current from contact 5 to contact 7. Using a standard lock-in technique[9,12,23] with low excitation frequency (see Methods section), we separate the second harmonic voltage component $V^{2f} \propto I^2$ from the first harmonic voltage response $V^{1f} \propto I$ (see Methods section). Measurements are carried out at room temperature as well as 77 K.

In order to prove the existence of an SHA, we measure the thermovoltage $V^{2f}$ by the Pt-$Ni_{45}Cu_{55}$ thermocouple as a function of an in-plane magnetic field, shown in Fig. 4a at room temperature. The second harmonic resistance, $R^{2f} = V^{2f}/I^2$, is characterized by four abrupt changes corresponding to the switching from P to AP configuration and vice versa. On the right y-axis the difference between the thermocouple ($T_{TC}$) and reference temperature ($T_0 = 300$ K) is plotted. The spin heat valve signal $R_S^{2f} = R_P^{2f} - R_{AP}^{2f}$ of $-0.04$ V $A^{-2}$ corresponds to a temperature difference of $-5$ m$K$. At 77K (Fig. 4b), the spin heat valve signal is $-0.06$ V $A^{-2}$ corresponding to a temperature change of $-12$ m$K$ between P and AP. The background thermal resistance, $R_b^{2f} = \frac{R_P^{2f} + R_{AP}^{2f}}{2}$, is lower at 77K (21.15 V $A^{-2}$) than at room temperature (29.13 V $A^{-2}$) due to the smaller heater resistance. Similar values are found for two other samples from the same batch (see Supplementary B).

In Fig. 4c we show the four-probe electrical resistance of the nanopillar at room temperature as a function of the external magnetic field measured using contacts 6 and 8 while a charge current flows from contact 5 to contact 7. A spin valve signal of $-80$ mΩ is observed on a background resistance of 2.27Ω. By using our three-dimensional finite element model (3D-FEM) to fit the spin vale signal, we obtain a spin polarization $P_\sigma$ of 0.52, typical of the bulk spin polarization for permalloy[12,23,24]. As a consistency check, the spin-dependent Seebeck[9,23] and Peltier effects[12] are also measured in the same device (see Supplementary C).

Fitting the measured spin heat valve signal of $-0.04 V\ A^{-2}$ to the spin heat diffusion model under the assumption of equal polarizations $P_\kappa$ and $P_\sigma$ (see Supplementary A and E), leads to a spin heat relaxation length $\lambda_{Q,Py}$ of 1 nm in Permalloy, which is a one-fifth of its spin relaxation length of 5 nm[25]. Taking the same scaling for the copper layer we obtain a $\lambda_{Q,Cu}$ of 70 nm as one- fifth of $\lambda_{s,Cu} = 350$ n$m$.[22] The fact that $\lambda_Q > t_N$ proves that inter-spin and electron-phonon inelastic scattering is surprisingly weak in nanopillar devices even at room temperature.

Most material-dependent transport parameters at 77 K can be found in the literature (see Supplementary Table 1). To fit the measured spin valve signal at 77 K of $-160$ mΩ (Fig. 4d), we require a slightly higher spin polarization $P_\sigma$ of 0.59, in agreement with earlier reports[24]. From the measured spin heat valve signal of $-0.06\ V\ A^{-2}$ and $P_\kappa = P_\sigma = 0.59$, we obtain a $\lambda_{Q,Cu}$ of 150 nm, more than two times longer than the $\lambda_{Q,Cu}$ at room temperature, demonstrating the reduced inelastic scattering.

From the 3D-FEM and the above experimental results we can now estimate the difference in the effective temperatures of the spin-up and the spin-down channels in the copper layer. We find $T_\uparrow - T_\downarrow = 120$ mK (at room temperature) and 350 mK (at 77K), up to 10% of the temperature bias of $4K$ across the nanopillar for a current of 2 mA through the heater.

In our modeling we do not take in to account electrical or heat interface resistances[25]. We would like to emphasize that those would not modify the extracted SHA (see Supplementary H). However, the fitted spin heat relaxation lengths will be affected. In Supplementary H we find a $\lambda_{Q,N}$ for copper of the order of 8 nm in the limit of a pure interface model.

The SHA is a unique concept that deserves more study. Our results indicate that the spin heat relaxation length in copper is close to the recently measured charge heat relaxation length[20,21]. Indeed, at higher temperatures the inelastic scattering is thought to be dominated by phonons and is not spin selective. We should therefore interpret the results not as a temperature difference of thermalized spin channels. The SHA is rather a measure of the difference between non-thermalized spin distributions that can be parameterized by the effective temperature parameter[19].

In summary, we measured the difference between the effective temperatures for spin-up and spin-down electrons in heat current-biased nanopillar spin valves. A spin dependence of the heat conductance on the magnetic configuration of multilayered current-in-plane-GMR devices[26-28] has been observed before, but involves neither spin accumulation nor a spin-dependent temperature. Modulating the heat conductance of the nanopillar by the magnetization configurations allows control of the flow of heat across the nanopillar, opening up possibilities for room temperature magnetic thermal switches. While optical pump and probe techniques and hot-electron transistors can access spin-dependent relaxation processes only at high energies, conventional transport experiments are limited to very low temperatures. The spin heat valve measurement, on the contrary, offers a unique possibility to estimate inelastic scattering length at the Fermi energy both at low and elevated temperatures.

## Methods

**Fabrication**

One optical lithography step followed by eleven electron-beam lithography (EBL) steps were employed to make the device. For each step, materials were e-beam evaporated except for the $Ni_{45}Cu_{55}$ alloy, which was sputtered so as to maintain the bulk stoichiometry. First, a 40-nm-thick Pt Joule heater was deposited on a thermally oxidized Si substrate. Then, the Pt-Constantan ($Ni_{45}Cu_{55}$) thermocouple was realized on top a 10 nm thick Au layer. Then an 8-nm-thick $Al_2O_3$ layer was deposited over the Pt-Joule heater and the thermocouple to electrically isolate the bottom contact of the nanopillar. The insulating layer prevents the pick-up of any charge related effects. Then, a Pt bottom contact (60 nm thick) was deposited on top of the heater and thermocouple. In the next step, the $Ni_{80}Fe_{20}$ (15)/Cu(15)/ $Ni_{80}Fe_{20}$ (15)/Au(10), where numbers between the parentheses are the thicknesses in nanometers, was deposited without breaking the vacuum of the deposition chamber. Cross-linked polymethyl methacrylate (PMMA) around the nanopillar prevents short circuiting between the bottom and the top contact (130-nm-thick Au).

**Measurements and modelling**

All measurements were done using a standard lock-in technique at low frequency (f < 20 Hz) such that steady-state condition is reached and at the same time capacitive coupling is prevented. Because a measured signal $V$ has both linear and non-linear contributions given as $V = I \cdot R^{1f} + I^2 \cdot R^{2f}$, we used a multiple lock-in measurement to distinguish the first harmonic resistance $R^{1f} = V^{1f}/I$ from the second harmonic resistance $R^{2f} = V^{2f}/I^2$. To fully characterize the samples, four different measurements were performed. First, in the spin valve measurements, the four-probe resistance of the nanopillar was measured as a function of magnetic field from which the bulk conductivity polarization ($P_\sigma$) was obtained. Then we measure the spin-dependent Seebeck and spin-dependent Peltier effect in the same device. From these measurements, the spin polarizations of the Seebeck ($P_S$) and Peltier coefficients ($P_\Pi$) are obtained. By using the 3D-FEM (see supplementary E) together with the extracted values for $P_\sigma$, $P_S$ and $P_\Pi$, we determine the spin heat relaxation length. Measurements were taken both at room and 77 K.

Acknowledgement:

We would like to acknowledge B. Wolfs, M. de Roosz and J.G. Holstein for technical assistance. This work is part of the research program of the Foundation for Fundamental Research on Matter (FOM) and supported by NanoLab NL, EU FP7 ICT Grant No. 251759 MACALO and the Zernike Institute for Advanced Materials.

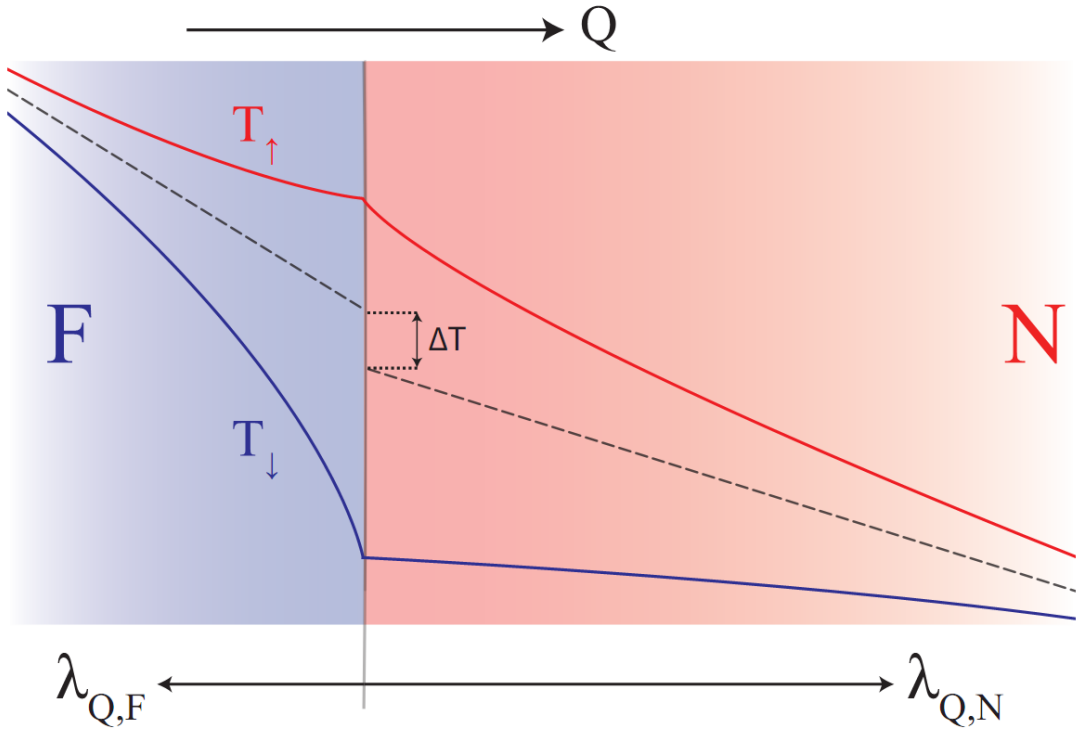

**Figure 1| Spin heat accumulation at an F/N interface.** The spin polarized heat current in a ferromagnetic metal (F) creates an SHA at the interface with a non-magnetic metal (N), because the heat currents have to be equally distributed over the spin channels in N. Inelastic scattering equilibrates the spin channel temperatures on the scale of the spin heat relaxation length $\lambda_Q$.

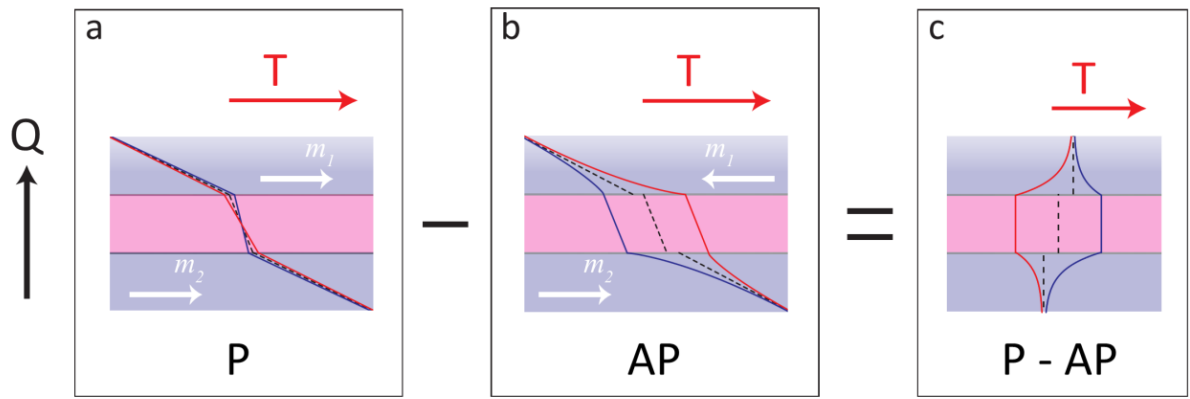

**Figure 2| Spin heat accumulations in an F/N/F spin valve.** Temperature profiles over the stack in the parallel (P) and antiparallel (AP) configuration in the presence of a heat current (Q) **a,** In the P configuration the SHA at both F/N interfaces have opposite signs, leading to a negligibly small SHA. **b,** For the AP configuration the SHA at the F/N interfaces have the same sign creating a large SHA and a corresponding temperature drop between the F/N interfaces and the bulk of the F layers. **c,** A temperature drop between the P and AP configuration builds up due to the spin heat valve effect.

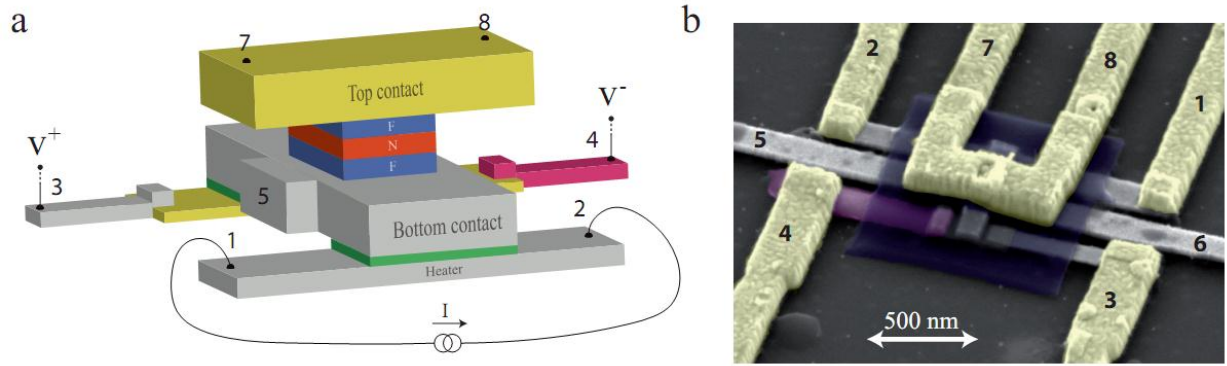

**Figure 3| Device geometry. a,** Schematics of the measured device showing an F/N/F pillar spin valve sandwiched between Au top and Pt bottom contacts. A charge current $I$ through the Pt-heater (contact 1 to 2) increases the temperature of the bottom contact, which is simultaneously measured by a Pt-Constantan ($Ni_{45}Cu_{55}$) thermocouple. Both the heater and thermocouple are electrically isolated from the bottom contact by an $Al_2O_3$ barrier (green; 8 nm thick) in order to avoid any charge-related spurious signals. **b,** Colored 3D-scanning electron microscope image of the measured device. The nanopillar sits halfway between the Pt-$Ni_{45}Cu_{55}$ thermocouple (contacts 3 and 4) and the Pt-Joule heater (contacts 1 and 2). Cross-linked PMMA (blue) electrically isolates the bottom contact (grey contacts 5 and 6) from the top contact (contacts 7 and 8).

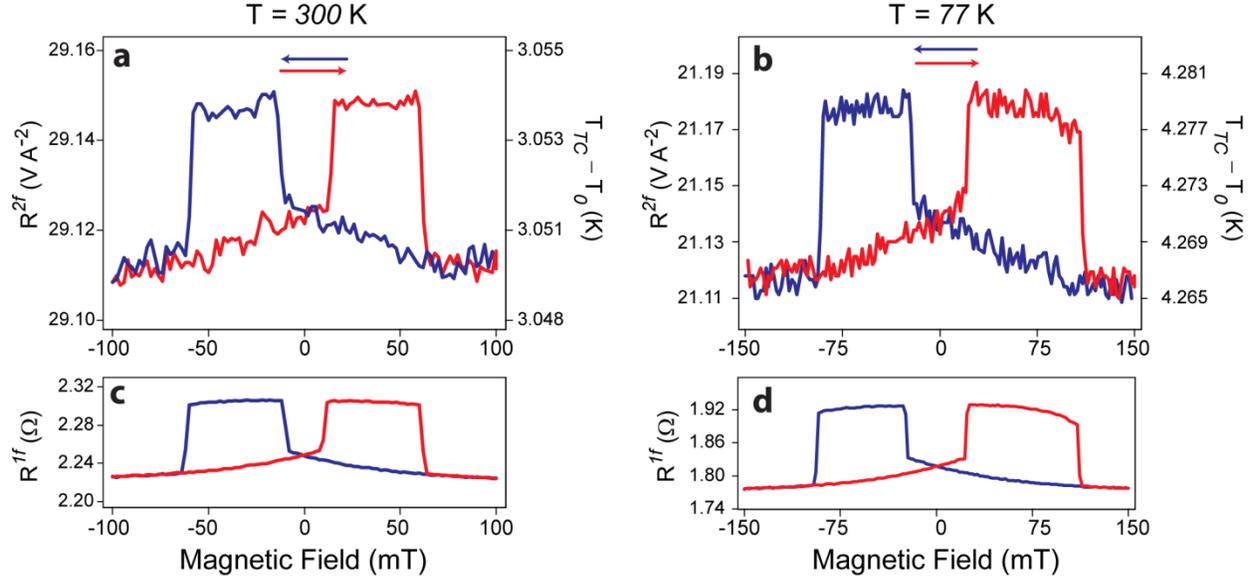

**Figure 4| Measured spin heat and conventional spin valve effects** Second harmonic response $R^{2f} = V^{2f}/I^2$ measured at the thermocouple **a,** at room temperature. **b,** at 77K, both for a current of 2 mA through the heater. Red and blue curves show forward ($-\vec{H} \rightarrow \vec{H}$) and backward ($\vec{H} \rightarrow -\vec{H}$) traces of the applied magnetic field. The right y-axis shows the temperature at the Pt-Ni$_{45}$Cu$_{55}$ thermocouple $T_{TC}$ relative to the reference room temperature $T_0 = 300$ K as $T_{TC} - T_0 = \frac{V^{2f}}{S_{NiCu} - S_{Pt}}$, where $S_{NiCu}$ and $S_{Pt}$ are the Seebeck coefficients for Ni$_{45}$Cu$_{55}$ and Pt (see Supplementary Table I). The heat resistance $R_{Q,pillar} \propto R^{2f}$ of the nanopillar and therefore the temperature is larger in the antiparallel than parallel configuration. **c,d,** Four-probe electrical resistances $R^{1f} = V^{1f}/I$ as a function of magnetic field measured using contacts 6 and 8 while current flows from contact 5 to contact 7 for room temperature and 77K, respectively.

# Supplementary Information

## A. Spin-dependent heat transport

The mathematical model for spin-dependent heat transport[3] is a thermal equivalent of the diffusion theory for spin-dependent charge transport[29] (for implementation of the model see supplementary E). As mentioned in the main text the spin polarization of the heat conductance in an F material leads to a spin heat accumulation (SHA) $T_s = T_\uparrow - T_\downarrow$ at an F/N interface in the presence of a heat current ($Q = -\kappa \cdot \nabla T$). This SHA obeys the diffusion equation

$$\nabla^2 T_s = \frac{T_s}{\lambda_Q^2} \tag{A1}$$

where $\lambda_Q$ is the material and temperature-dependent spin heat relaxation length. The solution to equation (A1) in the ferromagnetic metal reads

$$T_{\uparrow,\downarrow}(z) = A - \frac{Q}{\kappa_F} z \pm \frac{2B}{(1 \pm P_\kappa)\kappa_F} e^{+z/\lambda_{Q,F}} \tag{A2}$$

and in the normal metal

$$T_{\uparrow,\downarrow}(z) = \frac{Q}{\kappa_N} z \pm \frac{2C}{\kappa_N} e^{-z/\lambda_{Q,N}} \tag{A3}$$

with $P_\kappa = \frac{\kappa_{e\uparrow} - \kappa_{e\downarrow}}{\kappa_{e\uparrow} + \kappa_{e\downarrow}}$. The integration constants A, B and C are determined by the boundary conditions, namely continuity of $T_\uparrow$ and $T_\downarrow$ (in the absence of interface resistances) and conservation of heat currents $Q_\uparrow$ and $Q_\downarrow$ at the F/N interface. The SHA at the F/N interface then reads

$$T_s = \frac{2 \cdot P_\kappa \cdot \left(\frac{\lambda_{Q,F}}{\kappa_F}\right) \cdot \left(\frac{\lambda_{Q,N}}{\kappa_N}\right)}{\left(\frac{\lambda_{Q,F}}{\kappa_F}\right) + (1 - P_\kappa^2)\left(\frac{\lambda_{Q,N}}{\kappa_N}\right)} Q \tag{A4}$$

Interface heat resistances can significantly modify the $\lambda_{Q,F}$ and $\lambda_{Q,N}$ obtained from the bulk model as discussed in Supplementary section H.

At the F/N interface a spin-related thermal resistance leads to a temperature drop equal to the difference between the (particle) temperature in F, $T_F = \frac{\kappa_\uparrow T_\uparrow + \kappa_\downarrow T_\downarrow}{\kappa_F}$, and the temperature in N, $T_N = \frac{T_\uparrow + T_\downarrow}{2}$, giving a temperature difference of

$$\Delta T = \frac{1}{2} P_\kappa \cdot T_s \tag{A5}$$

In an F/N/F spin heat valve stack the temperature difference between the P and AP alignment of the magnetic layers over the entire pillar is $2 \cdot \Delta T$, assuming that $t_N \ll \lambda_{Q,N}$ with $t_N$ being the thickness of the N spacer.

## B. Results for two other samples

The spin heat valve effect was measured on two other samples fabricated in the same batch. Supplementary Fig. 1 shows the result of such measurements for sample 2 (Fig.1a and 1b) and sample 3 (Fig.1c and 1d), respectively, both at room temperature and 77K. At room temperature, both samples show a spin heat valve signals of $-0.04\, VA^{-2}$ on slightly different background signal of 30.97 V $A^{-2}$ and 28.87 $VA^{-2}$ respectively, similar to the sample presented in the main text. At 77K, sample 2 shows a slightly higher spin heat valve signal of $-0.07$ V $A^{-2}$ compared to sample 1. The spin heat valve signal of $-0.06$ V $A^{-2}$ in sample 3 is again similar to sample 1 in the main text. Also the spin heat relaxation length $\lambda_{Q,F}$ of 1 nm, $\lambda_{Q,N}$ of 50 nm (at room temperature) and 200 nm (at 77K) extracted from these measurements agree well with the values found for sample 1.

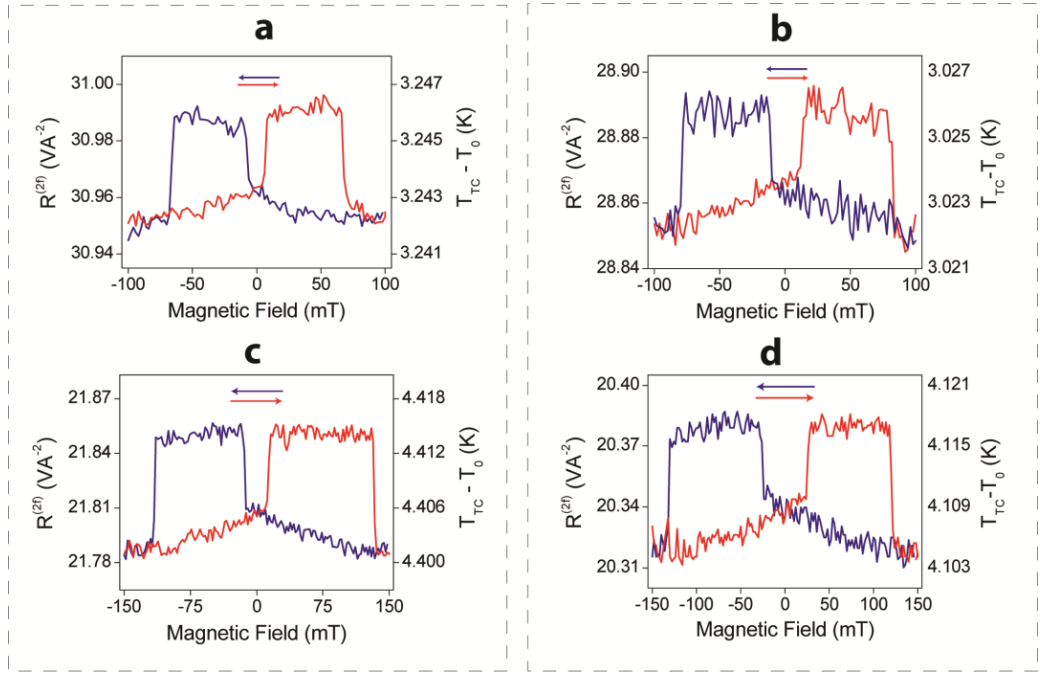

**Supplementary Figure 1| Spin heat valve measurement for two other samples.** The second harmonic resistance $R^{2f} = V^{2f}/I^2$ is plotted as a function of the applied magnetic field for sample 2 at room temperature (**a**) and at 77K (**b**) and for sample 3 at room temperature (**c**) and at 77K (**d**).

## C. Spin-dependent Peltier and spin-dependent Seebeck measurements

The spin-dependent Peltier effect[12] and the spin-dependent Seebeck effect[11,20] were also measured in samples 1, 2 and 3. Here we show the results for sample 1. From these measurements we obtain the spin polarization of the Seebeck and Peltier coefficient, which are later used in the modelling of the spin heat valve measurements. In the spin-dependent Seebeck effect[11,20], because of the difference in the Seebeck coefficients for spin-up ($S_\uparrow$) and spin-down ($S_\downarrow$) electrons, a temperature gradient $\nabla T$ across an F/N interface drives a spin current $J_S \propto (S_\uparrow - S_\downarrow)\nabla T$, where $S_s = S_\uparrow - S_\downarrow$ is the spin-dependent Seebeck coefficient of the ferromagnet, which is a fraction of the Seebeck coefficient of the ferromagnet $S_F$.

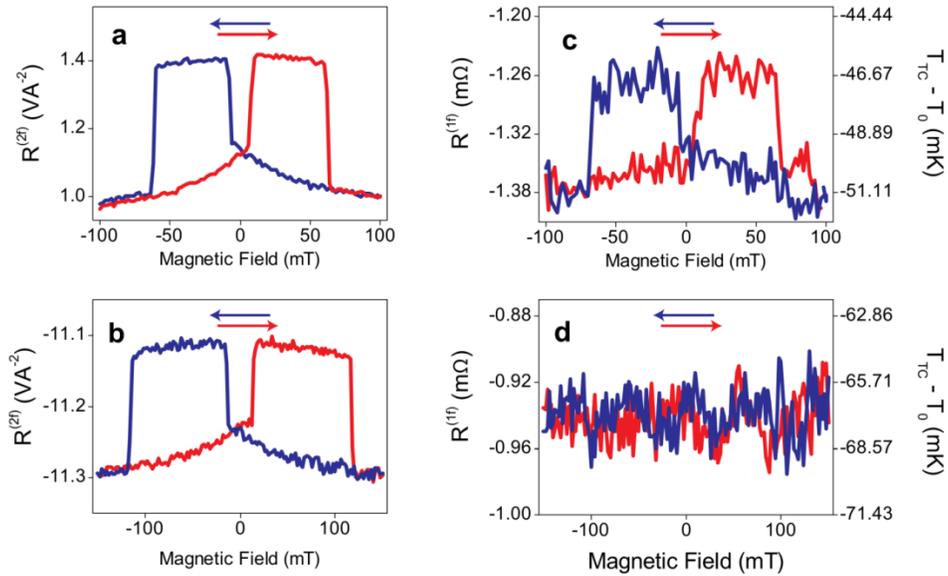

**Supplementary Figure 2| Spin-dependent Seebeck and spin-dependent Peltier effect measured for the sample presented in the main text.** The second harmonic signal $R^{2f} = \frac{V^{2f}}{I^2}$ as a function of the magnetic field in the spin-dependent Seebeck measurement (**a**) at room temperature and (**b**) at 77K, for a current of 2mA through the heater. The first harmonic signal $R^{1f} = \frac{V^{1f}}{I}$ as a function of the magnetic field in the spin-dependent Peltier measurement is also shown in **c** for room temperature and in **d** for 77K. The corresponding change in temperature is shown on the right y-axis.

In the spin-dependent Seebeck effect, we measure the open-circuit voltage across the nanopillar by using contacts 6 and 8 while sending a charge current of 2 mA through the Pt-heater. Supplementary Figs. 2a and 2b show the spin-dependent Seebeck signals for the sample 1 both at room temperature and at 77K, respectively.

From the plot of the second harmonic response $R^{2f} = V^{2f}/I^2$ as a function of the magnetic field we deduce a spin signal of $-0.4\ V\ A^{-2}$ and $-0.2\ V\ A^{-2}$ at room temperature and at 77K, respectively. The decrease in signal at lower temperatures is expected from the reduced Seebeck coefficient of Permalloy at such temperatures (see Supplementary Table 1). By fitting the finite element model to these signals, in which a small correction due to an SHA (see Supplementary D) is disregarded, we obtain a $P_S = \frac{S_\uparrow - S_\downarrow}{S_F}$ of $0.27 \pm 0.02$, which agrees with a previous report[9,23] of $P_S = 0.25$

The spin-dependent Peltier effect describes the heating/cooling of an F/N interface due to the flow of spin current $J_s$ through the interface. The temperature change is proportional to the spin accumulation $\mu_s^0$ at the interface and the difference of the Peltier coefficients for the spin-up and spin down electrons $\Pi_s = \Pi_\uparrow - \Pi_\downarrow$. In the experiment, a charge current is sent from contact 5 to 7 (Fig. 2b of the main text) and the thermovoltage is recorded using the thermocouple (contacts 2 and 3). Supplementary Figures 2c and 2d show the first harmonic response $R^{1f} = V^{1f}/I$ as a function of the magnetic field for room temperature and 77K, respectively. The corresponding temperature measured by the thermocouple is also plotted on the right y-axis. At room temperature, a spin-dependent Peltier signal of $-80\ \mu\Omega$ is observed on a background resistance of $-1.32$ m$\Omega$, in good agreement with earlier measurements in similar devices[14]. From the finite element model a spin-dependent Peltier coefficient of

−1.3 mV is obtained, demonstrating the Onsager-Kelvin relationship $\Pi_s = S_S T_0$ between the two spin-dependent thermoelectric transport coefficients. The low temperature results also confirm our earlier report that the spin-dependent Peltier signal vanishes quadratically with the temperature[12].

### D. Modification of $P_s$ by a spin heat accumulation

In our analysis of the spin-dependent Seebeck effect above and in our earlier reports[9, 23], we disregarded SHA ($\lambda_Q = 0$). The presence of an SHA modifies the spin polarization of the Seebeck coefficient $P_s$. By explicitly taking the SHA into account, we find that the previously determined value of $P_S = 0.25$ is increased to 0.35. This increase has, however, no significant impact on the analysis of the spin heat valve measurement as it contributes to the SHA only to higher order (see Supplementary G).

### E. Finite element modelling for spin and heat transport

We use a spin-dependent thermoelectric model in which the charge and heat currents in the two spin channels are defined in terms of the spin-dependent electrical conductivity $\sigma_{\uparrow,\downarrow} = \frac{\sigma}{2}(1 \pm P_\sigma)$, Seebeck coefficient $S_{\uparrow,\downarrow}$, Peltier coefficient $\Pi_{\uparrow,\downarrow} = S_{\uparrow,\downarrow} T_0$ for a reference temperature $T_0$, temperature $T_{\uparrow,\downarrow}$ and thermal conductivity $\kappa_{\uparrow,\downarrow} = \frac{\kappa}{2}(1 \pm P_\kappa)$, where $P_\kappa$ is the bulk spin polarization of the thermal conductivity. Here, $\kappa$ includes the contribution of the spin-dependent electronic thermal conductivity $\kappa_e$ and the phonon thermal conductivity $\kappa_{ph}$ that includes heat current paths through the insulating substrate. $P_\kappa$ is therefore a lower bound for the spin polarization of the electronic heat conductivity. The spin-dependent thermoelectric currents are then related to the driving forces as[3]

$$\begin{pmatrix} \vec{J}_\uparrow \\ \vec{J}_\downarrow \\ \vec{Q}_\uparrow \\ \vec{Q}_\downarrow \end{pmatrix} = - \begin{pmatrix} \sigma_\uparrow & 0 & \sigma_\uparrow S_\uparrow & 0 \\ 0 & \sigma_\downarrow & 0 & \sigma_\downarrow S_\downarrow \\ \sigma_\uparrow \Pi_\uparrow & 0 & \kappa_\uparrow & 0 \\ 0 & \sigma_\downarrow \Pi_\downarrow & 0 & \kappa_\downarrow \end{pmatrix} \begin{pmatrix} \vec{\nabla} V_\uparrow \\ \vec{\nabla} V_\downarrow \\ \vec{\nabla} T_\uparrow \\ \vec{\nabla} T_\downarrow \end{pmatrix} \quad (E1)$$

Spin relaxation due to spin-flip processes leads to non-conservation of spin currents while Joule heating causes finite divergence of the heat currents

$$\vec{\nabla} \cdot \begin{pmatrix} \vec{J}_\uparrow \\ \vec{J}_\downarrow \\ \vec{Q}_\uparrow \\ \vec{Q}_\downarrow \end{pmatrix} = \begin{pmatrix} \frac{(1-P_\sigma^2)\sigma}{4\lambda_s^2}(V_\uparrow - V_\downarrow) \\ -\frac{(1-P_\sigma^2)\sigma}{4\lambda_s^2}(V_\uparrow - V_\downarrow) \\ \frac{(1-P_\kappa^2)\kappa}{4\lambda_Q^2}(T_\uparrow - T_\downarrow) + \frac{J_\uparrow^2}{\sigma_\uparrow} + \frac{(1-P_\sigma^2)\sigma}{8\lambda_s^2}(V_\uparrow - V_\downarrow)^2 \\ -\frac{(1-P_\kappa^2)\kappa}{4\lambda_Q^2}(T_\uparrow - T_\downarrow) + \frac{J_\downarrow^2}{\sigma_\downarrow} + \frac{(1-P_\sigma^2)\sigma}{8\lambda_s^2}(V_\uparrow - V_\downarrow)^2 \end{pmatrix} \quad (E2)$$

The first two terms $\pm \frac{(1-P_\sigma^2)\sigma}{4\lambda_s^2}(V_\uparrow - V_\downarrow)$ represent spin relaxation between the spin up and spin down channels derived from the spin diffusion equation $\nabla^2 (V_\uparrow - V_\downarrow) = \frac{V_\uparrow - V_\downarrow}{\lambda_s^2}$. Joule heating in each channel is represented by $\frac{J_\uparrow^2}{\sigma_\uparrow}$ and $\frac{J_\downarrow^2}{\sigma_\downarrow}$. The term $\frac{(1-P_\sigma^2)\sigma}{4\lambda_s^2}(V_\uparrow - V_\downarrow)^2$ denotes heat generation due to spin relaxation[30]. Finally, the term $\pm \frac{(1-P_\kappa^2)\kappa}{2\lambda_Q^2}(T_\uparrow - T_\downarrow)$ denotes heat exchange between the two channels. Unlike the spin relaxation length, which is determined only by spin-flip scattering, the spin heat relaxation length $\lambda_Q$ is determined both by spin-flip scattering and inelastic scattering. The material parameters which are used in the model are listed in Supplementary Table 1. The room temperature data were separately measured and have previously been used in Ref. 23. Material parameters including Seebeck coefficients at 77 K are adopted from the literature. The spin relaxation lengths are taken from Ref. 25 and references therein.

| Material (thickness) | σ (RT) (10⁶ S m⁻¹) | S(RT) (μV K⁻¹) | κ(RT) (W m⁻¹ K⁻¹) | $\lambda_s$(RT) (nm) | σ(77K) (10⁶ S m⁻¹) | S (77K) (μV K⁻¹) | κ(77K) (W m⁻¹ K⁻¹) | $\lambda_s$(77K) (nm) |
|---|---|---|---|---|---|---|---|---|
| Au (130nm) | 27 | 1.7 | 180 | 80 | 40.5 [31] | 1.4 [32] | 107 [31] | 160 |
| Au (10nm) | 6.8 | 1.7 | 46 | 80 | 10 [31] | 1.4 [32] | 23 [31] | 160 |
| Pt (40nm) | 4.2 | -5 | 32 | 5 | 5.5 [33,36] | 6 [37] | 17 [30,33] | 10 |
| Pt (60nm) | 4.8 | -5 | 37 | 5 | 7.2 [33,34] | 6 [37] | 23 [33,36] | 10 |
| Cu (15nm) | 15 | 1.6 | 100 | 350 | 22.5 [35,36] | 1.3 | 60 [35,36] | 1000 |
| Py (15nm) | 2.9 | -18 | 18 | 5 | 4.3 [32] | -4.5 [32,34] | 11 [32,34] | 10 |
| Ni₄₅Cu₅₅ (30nm) | 2 | -32 | 20 | 5 | 3 | -8 [34,37] | 12 [34] | 10 |
| Al₂O₃ (8nm) | 0 | - | 0.12 | - | 0 | - | 0.1 | - |
| SiO₂ (300nm) | 0 | - | 1 | - | 0 | - | 0.1 | - |

**Supplementary Table 1: Material parameters used in the model.**

We strategically fit a single parameter per measurement in order to reduce large error margins caused by correlations between parameters. We first obtain the spin polarization $P_\sigma$ by fitting the measured electrical spin valve signal. By using this value in the analysis of the spin-dependent Seebeck and Peltier measurements, we obtain $P_S$ and $P_\Pi$, respectively. This allows us to accurately determine the spin polarization of the conductivity, Seebeck, and the Peltier coefficients. Having obtained these fitting results we model the spin heat valve measurement with the ratio of the spin heat relaxation length to the spin relaxation length $\frac{\lambda_Q}{\lambda_s}$ as a fitting parameter (see Supplementary F).

## F. Temperature profiles obtained from the finite element model

Here we show the temperature profile across an F/N/F perpendicular spin valve obtained using the 3D finite element model, for $\lambda_Q = 0.2\lambda_s$ (for both F and N) corresponding to the value obtained in the main text and for a charge current of 2mA through the Joule heater. In the parallel configuration, $T_\uparrow \approx T_\downarrow$ and any SHA is very small. In the antiparallel case, the spin heat accumulation in the copper layer is observed as a position-independent temperature difference of spin up and spin down electrons.

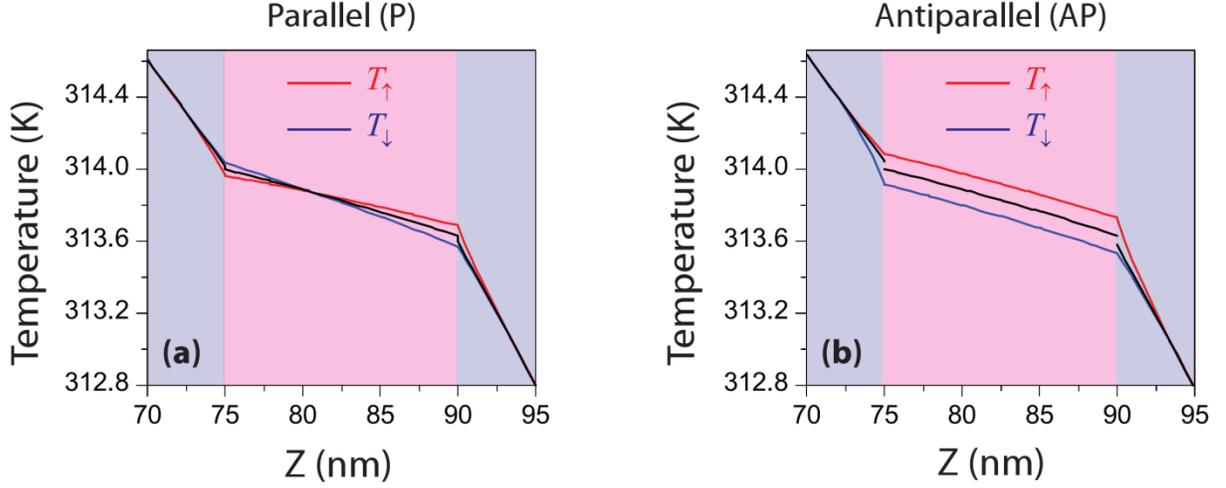

**Supplementary Figure 3| Temperature profile across the spin valve in the presence of spin-dependent temperatures for $\lambda_Q = 0.2\lambda_s$.** (a) In the parallel configuration, because the spin heat accumulation at the two F/N interfaces is opposite, the spin-dependent temperatures $T_{\uparrow,\downarrow}$ cross in the normal metal and the local temperature $T_0 = \frac{T_\uparrow + T_\downarrow}{2}$. (b) In the antiparallel configuration $T_{\uparrow,\downarrow} = T_0 \pm \frac{T_s}{2}$, where $T_s$ is the spin heat accumulation.

## G. <u>Possible contribution from spin-dependent Peltier effect induced by the spin-dependent Seebeck effect</u>

Here we discuss the interplay between the spin-dependent Seebeck and the spin-dependent Peltier effects that exists even in the absence of spin heat accumulation ($T_\uparrow = T_\downarrow$). When an F/N/F nanopillar stack is subjected to a temperature gradient, a thermally injected spin current from the first F/N interface (spin-dependent Seebeck effect) leads to heating/cooling of the second F/N interface (spin-dependent Peltier effect) and vice versa. By extensive model calculations, we find that being higher order in the thermoelectric coefficients this effect can contribute only ten percent to the measured spin heat valve signal at room temperature and is negligibly small at 77K, ruling out this effect as possible explanation for the heat valve effect.

## H. Modification of the spin heat relaxation lengths in a pure interface resistance model

In the main article and Supplementary E we only take into account bulk scattering. This is justified for charge transport in our Py/Cu/Py nanopillar because the contribution by the interface resistances is 4 times smaller than that of the bulk[25] and omission of the interfaces only leads to a slightly overestimated bulk polarization $P_\sigma$. The difference in heat resistance between P and AP alignment ($1/\Delta\kappa$) can be described in a similar way as $\Delta R$ for the GMR signal (Eq. (3) in ref. 25) by replacing the electrical resistivities and interface resistances by their thermal counterparts using the Wiedemann-Franz relation. Furthermore $\lambda_{S,Py}$ in the bulk term has to be replaced by its heat equivalent, $\lambda_{Q,Py}$, which is expected to be at least 5 times smaller (see the main text and Supplementary E) thereby reducing the bulk contribution to the spin heat accumulation. The interface contribution can therefore be of the same size or even dominate the spin heat signal, which requires a comparison of the extracted parameters from both analyses.

Here we explore the limit in which the spin-dependent thermal resistances are dominated by the interface. It is reasonable to assume that the interface resistances obey the Wiedemann-Franz relation and disregard any interface spin-flip scattering. In our current model this can be implemented by setting $P_\kappa = P_\sigma$ and $\lambda_{Q,Py} = \lambda_{S,Py}$ such that the SHA is limited by $\lambda_{Q,Cu}$ as is the case for a pure interface model. To fit the measured signal with this interface model we need to drastically decrease the spin heat relaxation length in copper ($\lambda_{S,Cu}$) to 7.5 nm (supplementary figure 4). It is important to notice that the interface spin heat accumulation does not change as it only depends on $P_\kappa$ and the observed heat-spin coupled $\Delta T$ (see equation 2).

Summarizing, we find that the extracted $T_\uparrow$, $T_\downarrow$ and $T_s$ are the same for both bulk and interface models (as well as intermediate regimes). However the spin heat relaxation lengths fitting the data significantly depend on the model assumptions about interface scattering.

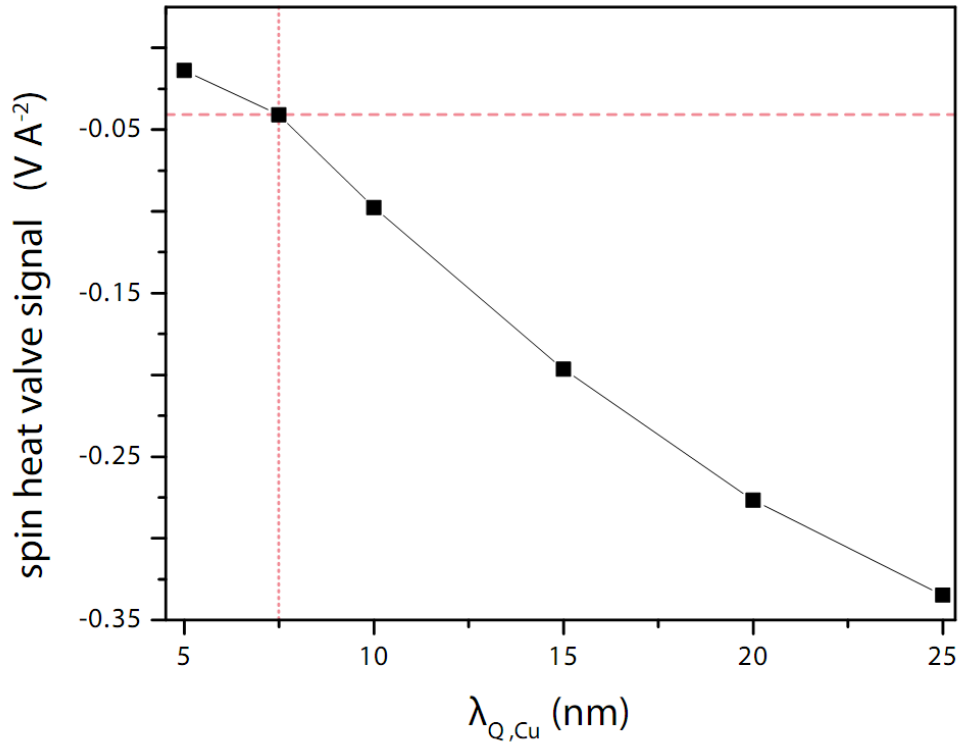

**Supplementary Figure 4| Spin heat valve signal versus the spin heat relaxation length in the copper ($\lambda_{S,Cu}$) for the pure interface resistance model.** The spin heat valve signal found by using a spin heat interface resistance model, by taking $P_\kappa = P_\sigma$ and $\lambda_{Q,Py} = \lambda_{S,Py}$. The measured spin signal (see main text and supplementary B) of -0.04 V A$^{-2}$ is indicated by the dashed line and the corresponding extracted $\lambda_{S,Cu}$ by the dotted line. The $\lambda_{S,Cu}$ of 7.5 nm is significantly smaller than the value derived from the model without interface resistance (see Supplementary E).